\documentclass[journal,twoside,web]{ieeecolor}
\usepackage{generic}
\usepackage{cite}
\usepackage{amsmath,amssymb,amsfonts}
\usepackage{algorithmic}
\usepackage{graphicx}
\usepackage{textcomp}
\usepackage{subfigure}
\usepackage{algorithm}
\newtheorem{theorem}{Theorem}

\newtheorem{corollary}{Corollary}

\newtheorem{lemma}{Lemma}
\newtheorem{remark}{Remark}
\newtheorem{definition}{Definition}
\newtheorem{assumption}{Assumption}

\definecolor{gray}{rgb}{0.5,0.5,0.5}

\newcommand{\pt}{\phi_k^{\mathrm{T}}\theta_k} 
\newcommand{\ptb}{\phi_k^{\mathrm{T}}\bar{\theta}_k}

\def\BibTeX{{\rm B\kern-.05em{\sc i\kern-.025em b}\kern-.08em
		T\kern-.1667em\lower.7ex\hbox{E}\kern-.125emX}}
\begin{document}
\title{$L_1$-Based Adaptive Identification with Saturated Observations}
\author{Xin Zheng and Lei Guo, \IEEEmembership{Fellow, IEEE}
    \thanks{This work was supported by the National Natural Science Foundation of China under Grant 12288201.}
    \thanks{Xin Zheng and Lei Guo are with the Key Laboratory of Systems and Control, Academy of Mathematics and Systems Science, Chinese Academy of Sciences, Beijing 100190, China, and also with the School of Mathematical Science, University of Chinese Academy of Sciences, Beijing 100049, China. (e-mails: zhengxin2021@amss.ac.cn, lguo@amss.ac.cn).}}

\maketitle	
\begin{abstract}

It is well-known that saturated output observations are prevalent in various practical systems and that the $\ell_1$-norm is more robust than the $\ell_2$-norm-based parameter estimation. Unfortunately, adaptive identification based on both saturated observations and the $\ell_1$-optimization turns out to be a challenging nonlinear problem, and has rarely been explored in the literature. Motivated by this and the need to fit with the $\ell_1$-based index of prediction accuracy in, e.g., judicial sentencing prediction problems, we propose a two-step weighted $\ell_1$-based adaptive identification algorithm. Under certain excitation conditions much weaker than the traditional persistent excitation (PE) condition, we will establish the global convergence of both the parameter estimators and the adaptive predictors. It is worth noting that our results do not rely on the widely used independent and identically distributed (iid) assumptions on the system signals, and thus do not exclude applications to feedback control systems.  We will demonstrate the advantages of our proposed new adaptive algorithm over the existing $\ell_2$-based ones, through both a numerical example and a real-data-based sentencing prediction problem. 	

\end{abstract}
	
	
\section{Introduction}\label{sec:introduction}

In various fields such as automatic control and signal processing, adaptive identification plays a pivotal role in revealing intrinsic system characteristics and optimizing control strategies based on collected observations. As a result, it has garnered considerable attention. However, in practical scenarios, most dynamical systems are often subject to various physical or other constraints, leading to the occurrence of saturated observations. Here, the saturated observation refers to a specific observation mechanism wherein the noise-corrupted output is accurately observable only within a certain range, and when the output surpasses this range, the observation becomes a constant value, resulting in imprecise information (\cite{zhang2023adaptive}). Moreover, the commonly used weighted $\ell_1$-norm for characterizing the accuracy of adaptive prediction exhibits many advantages over the widely used $\ell_2$-norm (\cite{Giloni2006}), making it a preferred choice for adaptive identification in the presence of outliers. Since the common occurrence of saturated observations and the advantages of weighted $\ell_1$-norm, adaptive identification based on such data and norm possesses significant theoretical and practical importance. 

Saturated observations in stochastic systems, emerge from various fields, including control systems (\cite{grip2009vehicle, sun2004aftertreatment, Yoneda2019automated}), communication (\cite{bandele2016saturation}), environmental monitoring (\cite{hanrahan2004electrochemical}), medical signal processing (\cite{lustig2007sparse, Craven2014Compressed}), economics (\cite{tobin1958estimation, jeon2020estimation, Bykhovskaya2023time}), and judiciary (\cite{wang2022applications}). To illustrate this point, we provide specific examples from three distinct fields. First, in the field of autonomous driving (\cite{Yoneda2019automated}), the observations from sensors such as cameras and LiDAR can be saturated due to the physical limitations of sensors and the presence of adverse weather conditions. Second, in the field of communication (\cite{bandele2016saturation}), cascaded optical amplifiers are employed to mitigate scintillation caused by atmospheric turbulence in free-space optical communication systems. The use of such optical amplifiers can result in saturated observations of the receivers in the systems. Third, in the field of judicial sentencing (\cite{wang2022applications}), the announced sentences of criminals are determined within a statutory range based on the criminal facts and relevant laws, and these announced sentences can be regarded as saturated observations. Compared with unsaturated observations, studying saturated observations poses highly non-trivial challenges due to the inherent non-linearity. 

Employing the weighted $\ell_1$-norm as a metric in describing the prediction accuracy index enhances the robustness of the estimator against outliers compared with both the unweighted $\ell_1$-norm and the usually used $\ell_2$-norm (\cite{Ellis1992leverage, Giloni2004, Giloni2006, wang2019robust, gao2018penalized}). For example, \cite{Ellis1992leverage} conducted research on small-sample datasets and found that the introduction of weighting coefficients into $\ell_1$-norm based least absolute deviation (LAD) regression could enhance the robustness of LAD regression, \cite{Giloni2004} examined this study in more detail. \cite{Giloni2006} demonstrated that a judicious choice of weights can enhance the robustness of weighted least absolute deviation (WLAD) regression compared with least squares (LS) regression and proved the convergence of the proposed estimator. Additionally, \cite{wang2019robust} utilized a more robust WLAD regression than the LS methods for derivative estimation in nonparametric regression, and derived the consistency and asymptotic normality of the estimator. Furthermore, \cite{gao2018penalized} proposed a penalized WLAD framework to improve the performance of LAD regression. Certainly, these studies are conducted on non-saturated linear models and depend on the assumption of iid (independent and identically distributed) signals or other stringent data conditions. Moreover, due to the inherent non-linearity of saturated observation models and the non-differentiability of the $\ell_1$-norm, there are many theoretical challenges associated with adaptive identification based on saturated observations and the weighted $\ell_1$-norm, and this direction has less been explored up to now.

Fortunately, there is significant progress on related problems of adaptive identification with saturated observations that can be referenced to help address these theoretical issues (\cite{powell1984least, heckman1976the, lai1992asymptotically, wang2007asymptotically, guo2013recursive}). For instance, \cite{powell1984least}  validated the strong consistency of the proposed estimators through the absolute deviation method under the assumptions that the regression vectors satisfy the iid conditions. \cite{lai1992asymptotically} established consistency and asymptotic efficiency of the estimators using the maximum likelihood method where the independent or non-random signals satisfying a stronger persistent excitation (PE) condition. \cite{wang2007asymptotically} established the strong consistency and asymptotic efficiency of the estimators using the empirical measure method under periodic signals with binary-valued observations. In fact, almost all of these existed results typically require the regression vectors to satisfy either iid conditions or conditions no weaker than PE. While this assumption makes the theoretical analysis easier, satisfying or verifying these conditions can be particularly challenging and even impossible, when dealing with real complex data, especially data from general dynamic systems with feedback loops (\cite{guo2020feedback}). Inspired by \cite{chen1991identification} and \cite{lai1982least} which studied adaptive identification with unsaturated observations and conditions of non-PE, \cite{zhang2022identification} established the strong consistency of estimators based on saturated observations without the need of PE conditions but under the usual $\ell_2$-norm optimization. Furthermore, \cite{zhang2023adaptive} extended this study to handle more general saturated observations and similarly established the strong consistency of the estimator and the asymptotic distribution of estimation errors based on non-PE conditions and $\ell_2$-norm optimization again. Since the weighted $\ell_1$-norm is more robust respect to outliers compared with the $\ell_2$-norm, and since the $\ell_1$-norm naturally correspond to the metric used in describing the prediction accuracy index, the design of the corresponding adaptive algorithms as well as the establishment of the convergence of the estimators under the saturated observations, are undoubtedly worthy of investigation.

To address the above-mentioned issues, we propose a new adaptive identification algorithm with general saturated observations based on the weighted $\ell_1$-norm optimization. Inspired by \cite{zhang2023adaptive}, we establish the global convergence of the proposed adaptive estimators and predictors under data conditions that are far weaker than the usual PE condition. The main contributions of this study are summarized as follows:
\begin{enumerate}
\item We will propose a new two-step weighted least absolute deviation (TSWLAD) adaptive identification algorithm tailored for saturated observations in stochastic dynamic systems based on weighted $\ell_1$-norm optimization. 

\item We will establish the global convergence of the new proposed algorithms based on a general non-PE condition. To the best of our knowledge, this is the first result in adaptive identification with saturated observations that establishes the global convergence of estimators based on the weighted $\ell_1$-norm optimization and non-PE conditions.

\item  We will also provide asymptotic upper bounds for the averaged regrets of adaptive predictors without requiring any PE conditions.
\end{enumerate}

The remainder of this paper is organized as follows. Section \ref{mainresults}  will present the main results, including the problem formulation, notations, assumptions, the proposed algorithms and main theorems. Section \ref{proofsmain} will provide the proofs for the main theorems. Section \ref{simulation} will demonstrate the advantages of the proposed algorithms with both simulation examples and sentencing problems with real data, and Section \ref{conclusion} will conclude this work with some remarks.

\section{The main results}\label{mainresults}
In this section, we will firstly outline the problem under investigation, then introduce the notations and assumptions employed in this paper, and finally present the newly proposed adaptive identification algorithm and theorems.

\subsection{Problem formulation}

We consider the following saturated stochastic model:
\begin{gather}\label{smodel}
	y_{k+1} = S_k(\phi_k^{\mathrm{T}}\theta+\varepsilon_{k+1}), \quad k = 0, 1, 2, \cdots,
\end{gather}
\noindent where $y_{k+1} \in \mathbb{R}$ represents the system observation, $\phi_k \in \mathbb{R}^d$ (with $d \geq 1$) denotes the stochastic regression vector, and $\varepsilon_{k+1} \in \mathbb{R}$ is the random noise. $\theta \in \mathbb{R}^d$ is the unknown parameter vector to be estimated, $S_k(\cdot): \mathbb{R} \rightarrow \mathbb{R}$ is a non-decreasing time-varying saturation function:
\begin{gather} \label{sfunction}
	S_k(x)=\left\{\begin{array}{cl}
	L_k, & x < l_k, \\
	x, & l_k \leq x \leq u_k , \\
	U_k, & x > u_k,
	\end{array}\right.
\end{gather}
\noindent where $\left[l_k, u_k\right]$ is the given observable range, $L_k$ and $U_k$ are the only observations when the system output surpasses this observable range.

Based on the online-accessed dataset $\left\{y_{i+1}, \phi_i\right\}_{i= 0}^{k}$ from the saturated stochastic model \eqref{smodel} at time $k$, one of our objectives is to design an adaptive predictor $\hat{y}_{k+1}$ that minimizes the following averaged weighted $\ell_1$-loss function, which is a metric in describing the prediction accuracy:
\begin{gather}\label{loss}
	\frac{1}{k+1}\sum\limits_{i= 0}^{k} b_{i} \vert y_{i+1} - \hat{y}_{i+1}\vert,
\end{gather}
where $b_i$ denotes a given bounded positive weighting coefficient which may be chosen according to the importance of the prediction error or to the need of regulation. Another critical objective is to achieve strongly consistent estimators of the unknown parameters $\theta$. To realize these objectives, we need the following notations and assumptions.

\subsection{Notations and assumptions}

\noindent\textbf{Notations.} We denote the Euclidean norm for vectors or matrices by $\|\cdot\|$; For a matrix $A$, $\lambda_{\max }\{A\}$ and $\lambda_{\min }\{A\}$ denote the maximum and minimum eigenvalues of $A$ respectively, $\operatorname{tr}(A)$ stands for the trace of $A$, and $|A|$ signifies the determinant of $A$; $\{\mathcal{F}_k\}_{k \geq 0}$ denotes the sequence of $\sigma$-algebras along with the associated conditional expectation $\mathbb{E}\left[\cdot \mid \mathcal{F}_k\right]$, the abbreviation $\mathbb{E}_k\left[\cdot\right]$ or $\mathbb{E}_k\left\{\cdot\right\}$ may sometimes be employed for $\mathbb{E}\left[\cdot \mid \mathcal{F}_k\right]$; The notation $\{x_k, \mathcal{F}_k\}_{k\geq 0}$  for a random sequence $\{x_k\}_{\geq 0}$ means that $x_k$ is $\mathcal{F}_k$-measurable for any $k\geq 0$; \(\operatorname{sgn}\left[\cdot\right]\) is the sign function and  $\operatorname{I}_{[\cdot]}$ is the indicator function.

We need the following assumptions:

\begin{assumption}\label{A1}
The stochastic regressor sequence $\{\phi_k, \mathcal{F}_k\}_{k\geq 0}$ is bounded in $k$, the true parameter $\theta$ is an interior point of a known convex compact set $D \subseteq \mathbb{R}^d$.
\end{assumption}

According to Assumption \ref{A1}, there exists a bounded random sequence $\{C_k\}_{k\geq 0}$ such that 
\begin{gather} \label{boundC}
   \sup_{x \in D} |\phi_k^{\mathrm{T}}x| \leq C_k.
\end{gather}

\begin{assumption}
The thresholds $\{l_k, \mathcal{F}_k\}_{k \geq 0}$, \( \{u_k, \mathcal{F}_k\}_{k \geq 0} \), \( \{L_k, \mathcal{F}_k\}_{k \geq 0} \), and \( \{U_k, \mathcal{F}_k\}_{k \geq 0} \) are known stochastic sequences. They satisfy the following conditions for any \( k \geq 0 \):
\begin{gather}\label{nondecreasing}
    L_k \leq l_k \leq u_k \leq U_k, \quad \text{a.s.},
\end{gather}
and there exists a positive constant $M$ such that
\begin{gather} \label{upperlower}
    \sup_{k \geq 0} \max\{l_k, -u_k\} \leq M < \infty, \text{a.s.},
\end{gather}
\noindent and the scenario where $L_k = l_k = u_k = U_k$ does not hold almost surely. In addition, the weighting sequence $\{b_k, \mathcal{F}_k\}_{k \geq 0}$ in \eqref{loss} is a known adapted sequence satisfying 
\begin{gather} \label{b_kbound}
	0<\inf_{k\geq 0} b_k \leq \sup_{k\geq 0}b_k \leq 1.
\end{gather}
\end{assumption} 

\begin{remark}
   Conditions \eqref{nondecreasing}  and \eqref{upperlower} are determined by the non-decreasing property of the saturation function \eqref{sfunction}. Condition \eqref{b_kbound} is a natural regularisation condition since optimizing the objective \eqref{loss} directly is equivalent to  optimizing the objective \eqref{loss} multiplied by $\frac{1}{M_0}$ where $M_0$ is an upper bound of $\{b_k\}_{k \geq 0}$.
\end{remark}

\begin{assumption}\label{symmetry}
The conditional distribution function of the noise $\varepsilon_{k+1}$ given the $\sigma$-algebra $\mathcal{F}_k$, denoted by $F_{k+1}(\cdot)$, satisfies $F_{k+1}(0) = \frac{1}{2}$. Moreover, the corresponding conditional density function $f_{k+1}(\cdot)$, is continuous and known. Furthermore, assume that there exists a constant $C$ such that $C \geq \sup\limits_{k\geq 0} C_k$ and that
\begin{gather}\label{boundednessf}
\begin{aligned}
    0 &< \inf_{|x| \leq \max\{2C, C+M\},  k \geq 0} f_{k+1}(x) \\
    &\leq \sup_{|x| \leq \max\{2C, C+M\}, k \geq 0} f_{k+1}(x) < \infty, \quad \text{a.s.}
\end{aligned}
\end{gather}
\end{assumption}

\begin{remark}
Assumption 3  needs the conditional distribution function and conditional density function of the noise  $\varepsilon_{k+1}$, along with the continuity of the conditional density function. These conditions facilitate the construction of the adaptive identification algorithm. As illustrated in sentencing problems (\cite{wang2022applications}), in practical applications, the distribution function may be estimated from real data through methods such as noise distribution fitting. Furthermore, the requirement that $ F_{k+1}(0) = \frac{1}{2}$ is just for simplicity in constructing the best predictor \eqref{loss}. In the case where $F_{k+1}(\zeta) = \frac{1}{2}$ but $\zeta \neq 0$, one may just do a simple modification for the best predictor to eliminate the impact of the mismatch of $\zeta$ with $\frac{1}{2}$. Details will be omitted here.
\end{remark}

\subsection{Adaptive identification algorithm}

To introduce a suitable adaptive predictor in \eqref{loss}, we need the following definition for the conditional median (\cite{Tomkins1975, chow1997}):
\begin{definition}\label{mediand}
Given a non-decreasing $\sigma$-algebra sequence $\{\mathcal{F}_k\}_{ k \geq 0}$ and an adapted sequence $\{x_k, \mathcal{F}_k\}_{k\geq 0}$, the conditional probability for the random variable $x_{k+1}$ given the $\sigma$-algebra $\mathcal{F}_k$ is denoted as $\mathbb{P}_{k+1}(\cdot)$. Therefore, the corresponding $\mathcal{F}_k$-measurable conditional median of $x_{k+1}$, denoted as $m_k$, satisfies $\mathbb{P}_{k+1}(x_{k+1} \leq m_k) \geq \frac{1}{2}$  and $\mathbb{P}_{k+1}(x_{k+1} \geq m_k) \geq \frac{1}{2}$.
\end{definition}

\begin{remark}
With the above definition, it is equivalent to saying that the conditional median $m_k$ satisfies $\mathbb{P}_{k+1}(x_{k+1} > m_k) \leq \frac{1}{2}$ and $\mathbb{P}_{k+1}(x_{k+1} < m_k) \leq \frac{1}{2}$.
\end{remark}

Moreover, the defined conditional median possesses the following property:
\begin{lemma} (\cite{Armerin2014}). \label{median} 
If $\mathbb{E}[|x_{k+1}| ] < \infty$, then the $\mathcal{F}_k$-measurable random variable $\rho$ that minimizes $\mathbb{E}[| x_{k+1} - \rho| ]$ is the conditional median of $x_{k+1}$ specified in Definition \ref{mediand}.
\end{lemma}   

\begin{remark}
In contrast to Lemma \ref{median}, it is well known that the minima that minimizes the $\ell_2$-norm based conditional mean square error is the conditional expectation. This fact makes our $\ell_1$-norm optimization based design of adaptive algorithms different fundamentally with the previous $\ell_2$-norm based algorithms (\cite{zhang2023adaptive}).
\end{remark}

With Lemma \ref{median} and Assumptions \ref{A1}--\ref{symmetry} in consideration, and for the purpose of parameter estimation, it is not difficult to deduce that the best predictor of (\ref{loss}) at time $k$ is as follows:
\begin{equation} \label{predict}
	\hat{y}_{k+1}^{*} =  S_k(\phi_k^{\mathrm{T}}\theta).
\end{equation}

Since the true parameter $\theta$ is unknown to us, we need to design an adaptive identification algorithm to estimate it at time $k$, the corresponding estimate is denoted by $\theta_k$. After obtaining the $\theta_k$, we replace the $\theta$ in \eqref{predict} with the $\theta_k$, resulting in an adaptive predictor as follows: 
\begin{gather}
\hat{y}_{k+1} = S_k(\phi_k^{\mathrm{T}}\theta_k). 
\end{gather}
 
For convenience of discussions, we will use the Frobenius norm $\|\cdot\|_Q$ associated with a positive definite matrix $Q$ such that $\|x\|_Q^2=$ $x^T Q x$ for any $x \in \mathbb{R}^d$. Furthermore, to ensure the estimates remain bounded and retain desirable properties during the computational process of the algorithm, we introduce a projection operator $\Pi_Q(\cdot)$ below (\cite{zhang2022identification}).

\begin{definition}\label{prodef}
The projection operator $\Pi_Q(\cdot)$ is defined as
\begin{equation}
	\Pi_Q(x)=\underset{y \in D}{\arg \min }\|x-y\|_Q, \quad \forall x \in \mathbb{R}^d ,
\end{equation}
where $D$ is the convex compact set defined in Assumption \ref{A1} and  $Q$ is a positive definite matrix.
\end{definition}

We are now poised to introduce the new adaptive identification algorithms motivated by the metric \eqref{loss}, the details of this algorithm are outlined in Algorithm \ref{tswla} below.

\begin{figure}[!ht]
\centering
\begin{algorithm}[H]
\caption{The Two-Step Weighted Least Absolute Deviation (TSWLAD) Algorithm}
\label{tswla}
\textbf{Step 1.} Recursively compute the preliminary estimates $\bar{\theta}_{k+1}$ for $k \geq 0$ :
\begin{gather}\label{algo11}
    \begin{aligned}
    \bar{\theta}_{k+1}=&\Pi_{\bar{P}_{k+1}^{-1}}\left\{\bar{\theta}_k+\bar{a}_k b_k \bar{P}_k \phi_k \bar{v}_{k+1}\right\},\\
    \bar{v}_{k+1} =& \operatorname{sgn}\left[y_{k+1}-S_k\left(\phi_k^{\mathrm{T}}\bar{\theta}_{k}\right)\right] \\
    & + F_{k+1}\left(u_k-\phi_k^{\mathrm{T}}\bar{\theta}_{k}\right)\operatorname{I}_{\left[S_k\left(\phi_k^{\mathrm{T}}\bar{\theta}_{k}\right) = U_k\right]}\\
    &-\left[1-F_{k+1}\left(l_k-\phi_k^{\mathrm{T}}\bar{\theta}_{k}\right)\right]\operatorname{I}_{\left[S_k\left(\phi_k^{\mathrm{T}}\bar{\theta}_{k}\right) = L_k\right]} 
     , \\
    \bar{P}_{k+1}  =&\bar{P}_k-\bar{a}_k \bar{\beta}_k b_k^2 \bar{P}_k \phi_k \phi_k^T \bar{P}_k, \\
    \bar{a}_k  =&\frac{1}{\bar{\mu}_k + \bar{\beta}_k b_k^2 \phi_k^T \bar{P}_k \phi_k}, \\
    \bar{\beta}_k =& \inf_{\parallel x \parallel \leq \max\{2C_k, C_k+l_k, C_k -u_k\}  }  f_{k+1}(x),
    \end{aligned}				
\end{gather}
where $\Pi_{\bar{P}_{k+1}^{-1}}^k\{\cdot\}$ is the projection operator defined in Definition \ref{prodef},  $\left\{\bar{\mu}_k, \mathcal{F}_k\right\}_{k \geq 0}$ is any positive sequence satisfying $0 < \inf\limits_{k\geq0} \bar{\mu}_k \leq \sup\limits_{k\geq 0} \bar{\mu}_k < \infty$. The initial values $\bar{\theta}_0$ and $\bar{P}_0$ can be randomly selected from the set $D$ and  with $\bar{P}_0 > 0$.
 
\textbf{Step 2.} Recursively compute the accelerated estimates $\theta_{k+1}$ for $k \geq 0$:
\begin{gather} \label{algo2}
    \begin{aligned}
    \theta_{k+1}=&\Pi_{P_{k+1}^{-1}}\left\{\theta_k+a_k b_k P_k \phi_k v_{k+1}\right\},\\
    v_{k+1} =& \operatorname{sgn}\left[y_{k+1} - S_k\left(\phi_k^{\mathrm{T}}\theta_{k}\right)\right]\\
    &+F_{k+1}\left(u_k-\phi_k^{\mathrm{T}}\theta_{k}\right)\operatorname{I}_{\left[S_k\left(\phi_k^{\mathrm{T}}\theta_{k}\right)= U_k\right]} \\
    &-\left[1 - F_{k+1}(l_k - \phi_k^{\mathrm{T}}\theta_{k})\right]\operatorname{I}_{\left[S_k\left(\phi_k^{\mathrm{T}}\theta_{k}\right)= L_k\right]} 
    , \\
    P_{k+1}  =&P_k-a_k \beta_k b_k^2 P_k \phi_k \phi_k^T P_k, \\
    a_k  =&\frac{1}{\mu_k + \beta_k b_k^2 \phi_k^T P_k \phi_k}, \\
    \beta_k =& \begin{cases}
        \frac{F_{k+1}(l_k - \phi_k^{\mathrm{T}}\theta_{k})-F_{k+1}(l_k-\phi_k^{\mathrm{T}}\bar{\theta}_{k})}{d_k} \operatorname{I}_{\left[d_k \neq 0\right]}\\
        + f_{k+1}(l_k - \phi_k^{\mathrm{T}}\theta_{k})\operatorname{I}_{[d_k =0]}, \ \text{if} \ S_k\left(\phi_k^{\mathrm{T}}\theta_{k}\right)= L_k, \\
        \\
        \frac{1-2F_{k+1}(\phi_k^{\mathrm{T}}\theta_{k}- \phi_k^{\mathrm{T}}\bar{\theta}_{k})}{d_k}\operatorname{I}_{\left[d_k\neq 0\right]}\\
        +2f_{k+1}(0)\operatorname{I}_{[d_k=0]},  \quad \text{if} \ L_k < S_k\left(\phi_k^{\mathrm{T}}\theta_{k}\right)  < U_k,\\
        \\
        \frac{F_{k+1}(u_k-\phi_k^{\mathrm{T}}\theta_{k})-F_{k+1}(u_k-\phi_k^{\mathrm{T}}\bar{\theta}_{k})}{d_k}\operatorname{I}_{[d_k \neq 0]} \\
        + f_{k+1}(u_k-\phi_k^{\mathrm{T}}\theta_{k})\operatorname{I}_{[d_k =0 ]}, \ \text{if}\ S_k\left(\phi_k^{\mathrm{T}}\theta_{k}\right) = U_k,
    \end{cases}
    \end{aligned}
\end{gather}
where $d_k = \phi_k^{\mathrm{T}}(\bar{\theta}_{k}-\theta_{k})$, $\left\{\mu_k\right\}_{k \geq 0}$ is defined similarly to $\left\{\bar{\mu}_k\right\}_{k \geq 0}$. The initial values $\theta_0$ and $P_0$  can be arbitrarily selected from the set $D$ and  with $P_0 > 0$.
\end{algorithm}
\vspace{-2em}
\end{figure}

We briefly explain how the TSWLAD algorithm is devised based on the least absolute deviation. Our focus is on clarifying the main ideas behind \eqref{algo11} only, since the primary difference between \eqref{algo11} and \eqref{algo2} lies in the adaptive gains \(\bar{\beta}_k\) and \(\beta_k\). The clarifications are presented in two steps:

\textbf{(1) The design of the gradient descent direction}. 

At time \(k\), based on the best predictor \eqref{predict}, the goal of our adaptive algorithm is to derive the next estimate of \(\theta\), denoted by \(\bar{\theta}_{k+1}\), by utilizing the data \(\{y_{k+1}, \phi_k\}\) from time \(k\) and the gradient direction of the  weighted absolute deviation objective function $b_k|y_{k+1} - S_k(\phi_k^{\mathrm{T}} \hat{\theta})|$ with respect to the estimate $\hat{\theta} \in \mathbb{R}^d$, evaluated at \(\hat{\theta} = \bar{\theta}_{k}\). This gradient can be expressed as $-b_k S_k'(\phi_k^{\mathrm{T}} \bar{\theta}_k) \phi_k \operatorname{sgn}[y_{k+1} - S_k(\phi_k^{\mathrm{T}} \bar{\theta}_k)]$  via simple calculations, where $S_k'(\cdot)$ denotes the derivative of \(S_k(\cdot)\), while neglecting the two non-differentiable zero-measure points.

\textbf{(2) Modification of the descent direction}.

In the previous descent direction, \( S_k'(\cdot) \) takes values of either 1 or 0. However, during computation, many regression vectors may cause \( \phi_k^{\mathrm{T}} \bar{\theta}_k \) to exceed the observable range \([l_k, u_k]\), resulting in \( S_k'(\phi_k^{\mathrm{T}} \bar{\theta}_k) = 0 \), which indicates that the parameters will not be updated. This situation not only results in the information from this data being unused but also complicates the theoretical analysis of parameter convergence. To address this issue, we substitute \( S_k'(\cdot) \) with 1. To ensure that the gradient is zero when the parameters converge to the true value, we modify the gradient by introducing two complement terms, see $\bar{v}_{k+1}$ in (12). This will ensure that the modified update direction satisfies \( \mathbb{E}[\bar{v}_{k+1}|\mathcal{F}_k] = 0 \) when the parameter estimate converges to its true value.

Next, we give a remark to clarify why the algorithm is designed in two steps.

\begin{remark}
Our new adaptive algorithm is a two-step procedure, with the main difference between the two steps of the new algorithm lying in the convergence rate of parameter estimation: the second step is faster than the first. This fact arises because, we select the adaptation gain $\bar{\beta}_k$ as the infimum of the density function specified in \eqref{algo11} to ensure the convergence of parameter estimation in the first step of the algorithm. However, the adaptation gain $\bar{\beta}_k$ may lead to slow convergence of parameter estimation since it may be extremely small, resulting in extremely slow updating of the $\bar{P}_k$ and $\bar{a}_k$. To address this issue, we firstly introduce the step-size adjustment factor $\bar{\mu}_k$ which may be chosen in various ways to adjust the convergence speed. However, the effectiveness of this adjustment may not be significant. Therefore, we design the second step of the algorithm to accelerate the convergence of estimation, where the new adaptive gain $\beta_k$ is determined based on the parameter estimate $\bar{\theta}_k$ from the previous step, and the step-size adjustment factor $\mu_k$ may also have appropriate parameter estimation convergence speed adjustment capabilities. Actually, the new adaptive gain \(\beta_k\) is crucial for accelerating the convergence of the estimates since it is comparatively larger, resulting in faster updating of the \(P_k\) and $a_k$. The above two-step procedure is motivated by \cite{zhang2023adaptive}, but the details of the two steps are different due to the difference of the metrics used for the prediction error.
\end{remark}

In addition, the choices of the hyperparameters of the algorithm are quite flexible, which may influence the performance of the algorithm: $C_k$ is the upper bound related to the regression vector in \eqref{boundC} and large $C_k$ will lead to small $\bar{\beta}_k$, and hence slow update rate in the gain matrix $\bar{P}_k$; $b_k$ is a given bounded positive weighting coefficient in \eqref{loss} which may be chosen according to the importance of the prediction error or to the need of regulation, and small $b_k$ will lead to slow adaptation gain; Both $\bar{\mu}_k$ and $ \mu_k $ are designed parameters that will influence the adaptation rate, since large values will give rise to slow adaptation rate.

\subsection{The main theorems}

In the following, we will establish the asymptotic upper bounds for both the parameter estimation errors and the averaged regrets of adaptive prediction.

\begin{theorem}\label{Th1}
Under Assumptions \ref{A1}--\ref{symmetry}, the estimate $\theta_{k+1}$ given by the TSWLAD algorithm has the following upper bound as $k \rightarrow \infty$: 
	\begin{gather}
		\|\tilde{\theta}_{k+1}\|^2 = O\left(\frac{\log \lambda_{\max}(k)}{\lambda_{\min}(k)}\right), \quad \text{a.s.}, 
	\end{gather}
\noindent where $\tilde{\theta}_{k+1} = \theta - \theta_{k+1}$, \(\lambda_{\max}(k)\) and \(\lambda_{\min}(k)\) denote the largest and smallest eigenvalues of the information matrix \(P_0^{-1} + \sum\limits_{i=0}^{k} \phi_i \phi_i^{\mathrm{T}}\), respectively.
\end{theorem}

The proof of Theorem \ref{Th1} is given in Section \ref{proofsmain}.

\begin{remark}
According to Theorem \ref{Th1}, the estimates $\{\theta_k\}_{k \geq 0}$ will converge to the true parameter $\theta$ almost surely if
\begin{equation}\label{npe}
    \lim_{k \rightarrow \infty} \frac{\log \lambda_{\max}(k)}{\lambda_{\min}(k)} = 0, \quad \text{a.s.}
\end{equation} 
\noindent The condition \eqref{npe} does not need the independence and stationarity of the regression vectors and is therefore applicable to stochastic feedback control systems. This condition is far weaker than the traditional PE condition (i.e., $\lambda_{max}(k)=O(\lambda_{min}(k))$), and coincide with the excitation condition derived under $\ell_2$-norm optimization for saturated observation models (\cite{zhang2023adaptive}), and is known as the weakest possible excitation condition for the well-known least squares estimates of the standard linear regression model (\cite{lai1982least}). Note that, the global convergence of the parameter estimates ensures that the projection operator will be unnecessary after some finite time since the true parameter is an interior of the compact set $D$.
\end{remark}

\begin{corollary} \label{corollary1}
Assuming the conditions outlined in Theorem \ref{Th1} are satisfied, and if the condition (\ref{npe}) is enhanced to $k = O(\lambda_{\min}(k))$, then the convergence rate of the TSWLAD algorithm can be improved as follows:
\begin{gather} \label{improvedrate}
    \|\tilde{\theta}_{k+1}\| = O(\frac{\log \log k}{k}), \quad \text{a.s.}
\end{gather}
\end{corollary}

The proof of Corollary \ref{corollary1} follows a similar way to that of Corollary 1 in \cite{zhang2023adaptive} and is therefore omitted here.

Furthermore, to measure the deviation between the adaptive predictor $\hat{y}_{k+1}$ and the optimal predictor $\hat{y}_{k+1}^{*}$, we define the corresponding regret as follows:
\begin{equation}
	R_k = b_k|\hat{y}_{k+1}^{*} - \hat{y}_{k+1}|.
\end{equation}
\noindent The next theorem provides an asymptotic upper bound for the averaged regrets.
\begin{theorem}\label{Th2}
Under Assumptions \ref{A1}--\ref{symmetry}, if the saturation function \eqref{sfunction} is continuous (i.e., $l_k =L_k$ and $u_k=U_k$), then the averaged regrets possess the following property:
\begin{gather}\label{regrettheorem}
    \frac{1}{n}\sum_{k=1}^{n}R_k = O\left(\sqrt{\frac{\log\lambda_{\max}(n)}{n}}\right), \quad \text{a.s.}
\end{gather}
\end{theorem} 

The proof of Theorem \ref{Th2} is given in Section \ref{proofsmain}. 

This result \eqref{regrettheorem} does not rely on any PE conditions, making it applicable to closed-loop control systems. Additionally, the continuous condition of the saturation function \eqref{sfunction} is quite general and satisfied in various practical systems, such as in judicial sentencing field (\cite{wang2022applications}).

As for the prediction error between the adaptive predictor $\hat{y}_{k+1}$ and the real observation $y_{k+1}$, we can obtain the corresponding upper bounds in the next theorem.

\begin{theorem}\label{Th3}
    Under the same conditions of Theorem \ref{Th2}, and assuming that the noise sequence $\{\varepsilon_{k}, \mathcal{F}_k\}_{k \geq 0}$ is a martingale difference sequence, we further assume the existence of a constant $r>4$ such that 
    \begin{gather}\label{noisemarting}
       \sup\limits_{k \geq 0} \mathbb{E}_k[|\varepsilon_{k+1}|^{r}] < \infty, \quad \text{a.s.}
    \end{gather}
    Then the averaged prediction error $E_n$ defined as
    \begin{gather}
        E_n = \frac{1}{n}\sum_{k=0}^{n-1}b_k|y_{k+1}- \hat{y}_{k+1}|,
    \end{gather}
    will have the following upper bound:
    \begin{gather}
    E_n \leq \sqrt{\frac{\sum\limits_{k=0}^{n-1}b_k^2\sigma_k^2}{n}} + O\left(  \sqrt[4]{\frac{\log\log n}{n}}  \right)  ,\quad \text{a.s.},
    \end{gather}
    \noindent where $\sigma_k^2 = \mathbb{E}_k\left[\varepsilon_{k+1}^2\right]$.
\end{theorem}

The proof of Theorem \ref{Th3} is in Section \ref{proofsmain}.

\section{Proofs of the main theorems \label{proofsmain}}

In this section, we will provide the proofs of the main theorems presented in this paper. For convenience in analysis, the following notations are introduced:
\begin{gather}\label{barpsik}
\begin{aligned}
\bar{\psi}_k =& \mathbb{E}_k\left\{\operatorname{sgn}\left[y_{k+1}-S_k(\ptb)\right] \right\} \\
&+F_{k+1}(u_k - \ptb)\operatorname{I}_{\left[S_k\left(\phi_k^{\mathrm{T}}\bar{\theta}_{k}\right) = U_k\right]}\\
&- \left[1- F_{k+1}(l_k - \ptb)\right]\operatorname{I}_{\left[S_k\left(\phi_k^{\mathrm{T}}\bar{\theta}_{k}\right) = L_k\right]},
\end{aligned}
\end{gather}
\begin{gather}\label{psik}
\begin{aligned}
\psi_k =& \mathbb{E}_k\left\{\operatorname{sgn}\left[y_{k+1}-S_k(\pt)\right] \right\} \\
&+F_{k+1}(u_k - \pt)\operatorname{I}_{[S_k\left(\phi_k^{\mathrm{T}}\theta_{k}\right) = U_k]}  \\
&-\left[1- F_{k+1}(l_k - \pt)\right]\operatorname{I}_{[S_k\left(\phi_k^{\mathrm{T}}\theta_{k}\right) = L_k]},
\end{aligned}
\end{gather}
besides, 
\begin{gather}
\begin{aligned}
 \bar{w}_{k+1} =& \operatorname{sgn} \left[y_{k+1} - S_k(\phi_k^{\mathrm{T}}\bar{\theta}_k)\right]\\
 &- \mathbb{E}_k\left\{ \operatorname{sgn}\left[y_{k+1}  - S_k(\phi_k^{\mathrm{T}}\bar{\theta}_k)\right]\right\},
 \end{aligned}
\end{gather}
\begin{equation}\label{wk+1}
\begin{aligned}
w_{k+1} = & \operatorname{sgn}\left[y_{k+1} - S_k(\pt)\right] \\
&-\mathbb{E}_k\left\{\operatorname{sgn}\left[y_{k+1} - S_k(\pt)\right]\right\}.
\end{aligned}
\end{equation}

With Assumptions \ref{A1}--\ref{symmetry}, we can deduce that  $\{\bar{w}_{k}, \mathcal{F}_k\}_{k \geq 0}$ and $\{w_{k}, \mathcal{F}_k\}_{k \geq 0}$ form  martingale difference sequences, satisfying $\sup\limits_{k\geq 0}\mathbb{E}_k\left[|\bar{w}_{k+1}|^{2+\eta}\right] < \infty$ and $\sup\limits_{k\geq 0}\mathbb{E}_k\left[|w_{k+1}|^{2+\eta}\right] < \infty$, where the constant $\eta>0$.

To prove Theorem \ref{Th1}, Theorem \ref{Th2} and Theorem \ref{Th3}, we need the following lemmas.

\begin{lemma}\label{project} (\cite{cheney2001}). 	
The projection operator given by Definition \ref{prodef} satisfies
\begin{equation}
\left\|\Pi_Q(x)-\Pi_Q(y)\right\|_Q \leq\|x-y\|_Q, ~ \forall x, y \in \mathbb{R}^d.
 \end{equation}
\end{lemma} 

\begin{lemma}\label{marting} (\cite{chen1991identification}). Let $\left\{w_n, \mathcal{F}_n\right\}_{n \geq 0}$ be a martingale difference sequence and $\left\{g_n, \mathcal{F}_n\right\}_{n \geq 0}$ an adapted sequence. If $\sup\limits_{n\geq 0} \mathbb{E}\left[\left|w_{n+1}\right|^\alpha \mid \mathcal{F}_n\right]<\infty, \text { a.s. }$, for some $\alpha \in(0,2]$, then as $n \rightarrow \infty$ :
\begin{gather}
\sum_{i=0}^n g_i w_{i+1}=O\left(s_n(\alpha) \log ^{\frac{1}{\alpha}+\eta}\left(s_n^\alpha(\alpha)+e\right)\right),~\text{a.s.}, \forall \eta>0,
\end{gather}
\noindent where $s_n(\alpha)=\left(\sum\limits_{i=0}^n\left|g_i\right|^\alpha\right)^{\frac{1}{\alpha}}.$
\end{lemma}

\begin{lemma}\label{lem1} 
(\cite{lai1982least}). Let $X_1, X_2, \cdots$ be a sequence of vectors in $\mathbb{R}^d(d \geq 1)$ and let $A_n=A_0+\sum\limits_{i=1}^n X_i X_i^\mathrm{T}$. Assume that $A_0$ is nonsingular, then as $n \rightarrow \infty$:
\begin{equation}
\sum_{k=1}^n \frac{X_k^T A_{k-1}^{-1} X_k}{1+X_k^T A_{k-1}^{-1} X_k} \leq \log \left(\left|A_n\right|\right)+\log \left(\left|A_0\right|\right) .
\end{equation}
\end{lemma}

\begin{lemma}\label{lem2}
(\cite{guo1995convergence}). Let $X_1, X_2, \cdots$ be any bounded sequence of vectors in $\mathbb{R}^d (d \geq 1)$. Denote $A_n=A_0+$ $\sum\limits_{i=1}^n X_i X_i^T$ with the matrix $A_0>0$, then we have
\begin{gather}	
\sum_{k=1}^{\infty}\left(X_k^T A_{k-1}^{-1} X_k\right)^2<\infty.
\end{gather}
\end{lemma}

\begin{lemma}
\label{l1}
Under Assumptions \ref{A1}--\ref{symmetry}, $|\bar{\psi}_k| \leq 1$ and $|\psi_k| \leq 1$. 
\end{lemma}

\begin{lemma} \label{critical}
Under Assumptions \ref{A1}--\ref{symmetry},	$|\psi_k - \beta_k \phi_k^{\mathrm{T}}\tilde{\theta}_k| =  O\left(|\phi_k^{\mathrm{T}}\tilde{\bar{\theta}}_k|\right),$ \text{a.s.}, where $\tilde{\bar{\theta}}_k =  \theta - \bar{\theta}_k$.
\end{lemma}

\begin{lemma}\label{lemma3}
Under Assumptions \ref{A1}--\ref{symmetry},  $|\bar{\psi}_k| \geq \bar{\beta}_k |\phi_k^{\mathrm{T}}\tilde{\bar{\theta}}_k|$ and $\phi_k^{\mathrm{T}}\tilde{\bar{\theta}}_k\bar{\psi}_k\geq 0$.
\end{lemma}

The proofs of Lemma \ref{l1}, Lemma \ref{critical} and \ref{lemma3} are in Appendix \ref{app1}.

\begin{lemma}\label{lemma1}
Let Assumptions \ref{A1}--\ref{symmetry} be satisfied, then the parameter estimate $\theta_{k+1}$ given by the TSWLAD algorithm  has the following property almost surely as $k \rightarrow \infty$:
	\begin{gather} \label{desired}
        \begin{aligned}
       \tilde{\theta}_{k+1}^T P_{k+1}^{-1} \tilde{\theta}_{k+1}+ \sum_{i=0}^k \mu_i^{-1}\beta_i^{-1}b_i \psi_i^2=O\left(\log \lambda_{\max }\left(k\right)\right)&,
       \end{aligned}
    \end{gather}
where $\tilde{\theta}_{k+1} = \theta-\theta_{k+1}$.
\end{lemma}

\noindent\hspace{1em}{\textbf{\itshape Proof of Lemma \ref{lemma1}:}} 
Inspired by the classical recursive least square algorithm for linear stochastic models in \cite{lai1982least}, \cite{guo1995convergence}, and the $\ell_2$-norm based methods for addressing the saturated-observations in \cite{zhang2023adaptive}, we choose the stochastic Lyapunov function as follows:
\begin{gather}\label{Vk}
    V_{k+1} = \tilde{\theta}_{k+1}^{\mathrm{T}}P_{k+1}^{-1}\tilde{\theta}_{k+1}.
\end{gather}
  
According to \eqref{algo2} and the well-known matrix inversion formula (see e.g., \cite{guo2020tima}), we have that
\begin{gather} \label{Pni}
    P_{k+1}^{-1} = P_{k}^{-1} + \mu_k^{-1}\beta_k b_k^2 \phi_k \phi_k^{\mathrm{T}}.
\end{gather}
Therefore, multiplying the left side of (\ref{Pni}) by $a_k\phi_k^{\mathrm{T}}P_k$ and taking into account the definition of $a_k$, we have
    \begin{equation} \label{leftP}
		\begin{aligned}
		    &a_k\phi_k^{\mathrm{T}}P_kP_{k+1}^{-1} \\
            =&a_k\phi_k^{\mathrm{T}} (I + \mu_k^{-1}\beta_k b_k^2 P_k\phi_k \phi_k^{\mathrm{T}}) = \mu_k^{-1} \phi_k^{\mathrm{T}}.
		\end{aligned}
	\end{equation}
Further, by Lemma \ref{project},  \eqref{Vk}) and \eqref{leftP} , we know that 
\begin{gather}\label{Vk+1}
    \begin{aligned}
		V_{k+1} \leq& \left[\tilde{\theta}_k - a_kb_kP_k\phi_k\left(\psi_k + w_{k+1}\right)\right]^{\mathrm{T}}P_{k+1}^{-1}\mathbf{\cdot}\\
          &\left[\tilde{\theta}_k - a_kb_kP_k\phi_k\left(\psi_k + w_{k+1}\right)\right]\\
        \leq& \tilde{\theta}_{k}^{\mathrm{T}}P_{k+1}^{-1}\tilde{\theta}_{k} - 2a_kb_k\phi_k^{\mathrm{T}}P_kP_{k+1}^{-1}\tilde{\theta}_{k}\psi_k \\
	   &+a_k^2 b_k^2 \phi_k^{\mathrm{T}} P_kP_{k+1}^{-1}P_k\phi_k\psi_k^2 \\
	   & +2a_k^2b_k^2 \phi_k^{\mathrm{T}} P_kP_{k+1}^{-1}P_k\phi_k \psi_kw_{k+1}\\
	   &-2a_kb_k\phi_k^{\mathrm{T}}P_kP_{k+1}^{-1}\tilde{\theta}_kw_{k+1} \\
	   & +a_k^2 b_k^2\phi_k^{\mathrm{T}}P_k P_{k+1}^{-1}P_k\phi_kw_{k+1}^2 \\
	   = & V_k +\mu_k^{-1} \beta_k b_k^2(\phi_k^{\mathrm{T}}\tilde{\theta}_k)^2 - 2\mu_k^{-1}b_k\phi_k^{\mathrm{T}}\tilde{\theta}_k \psi_k \\
	   &+\mu_k^{-1} a_k b_k^2\phi_k^{\mathrm{T}}P_k\phi_k \psi_k^2 \\
          & + 2\mu_k^{-1}a_kb_k^2 \phi_k^{\mathrm{T}}P_k\phi_k \psi_k w_{k+1} \\
	   & -2\mu_k^{-1}b_k\phi_k^{\mathrm{T}}\tilde{\theta}_kw_{k+1} + \mu_k^{-1}a_kb_k^2\phi_k^{\mathrm{T}}P_k\phi_kw_{k+1}^2, \\
    \end{aligned}
\end{gather}
where $\psi_k$ and $w_{k+1}$ are defined in \eqref{psik} and \eqref{wk+1}, respectively. Noticing that $0 < b_k \leq 1$, we have 
\begin{gather} \label{square1}
	\begin{aligned}
		&\mu_k^{-1} \beta_k b_k^2(\phi_k^{\mathrm{T}}\tilde{\theta}_k)^2 - 2\mu_k^{-1}b_k\phi_k^{\mathrm{T}}\tilde{\theta}_k \psi_k \\
	   \leq & \mu_k^{-1} \beta_k b_k(\phi_k^{\mathrm{T}}\tilde{\theta}_k)^2 - 2\mu_k^{-1}b_k\phi_k^{\mathrm{T}}\tilde{\theta}_k \psi_k\\
	   \leq & \mu_k^{-1}\beta_k^{-1}b_k( \psi_k - \beta_k\phi_k^{\mathrm{T}}\tilde{\theta}_k)^2 - \mu_k^{-1} \beta_k^{-1}b_k \psi_k^2, 
	\end{aligned}
\end{gather}
besides, 
\begin{gather}\label{square2}
    \begin{aligned}
	&\mu_k^{-1}a_kb_k^2 \phi_k^{\mathrm{T}}P_k\phi_k \psi_k w_{k+1} - \mu_k^{-1}b_k\phi_k^{\mathrm{T}}\tilde{\theta}_kw_{k+1} \\
	=& \mu_k^{-1}\beta_k^{-1}(1-\mu_ka_k)\psi_kw_{k+1} - \mu_k^{-1}b_k\phi_k^{\mathrm{T}}\tilde{\theta}_kw_{k+1} \\
	=& \mu_k^{-1} \beta_k^{-1}b_k(\psi_k - \beta_k\phi_k^{\mathrm{T}}\tilde{\theta}_k)w_{k+1}\\
         &+ \mu_k^{-1} \beta_k^{-1}(1-b_k-\mu_ka_k)\psi_k w_{k+1}.
    \end{aligned}
\end{gather}

Summing up both sides of \eqref{Vk+1} from 0 to $n$ and applying Lemma \ref{l1}, along with \eqref{square1} and \eqref{square2}, yields
\begin{gather} \label{Vc}
\begin{aligned}
    V_{n+1} \leq& V_0 - \sum_{k=0}^{n}\mu_k^{-1} \beta_k^{-1}b_k \psi_k^2   \\ 
    &+ \sum_{k=0}^{n}\mu_k^{-1}\beta_k^{-1}b_k( \psi_k - \beta_k\phi_k^{\mathrm{T}}\tilde{\theta}_k)^2  \\
    & + \frac{1}{h}\sum_{k=0}^{n}\mu_k^{-1}a_k\beta_kb_k^{2}\phi_k^{\mathrm{T}}P_k\phi_k  \\
    &+ 2\sum_{k=0}^{n}\mu_k^{-1}\beta_k^{-1}b_k(\psi_k - \beta_k\phi_k^{\mathrm{T}}\tilde{\theta}_k)w_{k+1}  \\
    &+2\sum_{k=0}^{n}\mu_k^{-1}\beta_k^{-1}(1-b_k-\mu_ka_k)\psi_kw_{k+1} \\
    &+ \frac{1}{h}\sum_{k=0}^{n}\mu_k^{-1}a_k\beta_k b_k^2\phi_k^{\mathrm{T}}P_k\phi_kw_{k+1}^2,\quad \text{a.s.} 
\end{aligned}
\end{gather}
\noindent where $h= \inf\limits_{|x|\leq \max\{2C, C+M\},\ k\geq 0} f_{k+1}(x)>0$.

Let us analyze the RHS of \eqref{Vc} term by term.

Firstly, for the fourth term on the RHS of \eqref{Vc}, since $0 < \inf\limits_{k\geq0} \mu_k \leq \sup\limits_{k\geq 0} \mu_k < \infty$, and let $X_{k}= \sqrt{\mu_{k}^{-1}\beta_k}b_k\phi_k$ in Lemma \ref{lem1}, we have
\begin{gather}\label{step1begin}
	\begin{aligned}
		&\sum_{k=0}^{n}\mu_k^{-1}a_k\beta_kb_k^{2}\phi_k^{\mathrm{T}}P_k\phi_k	\\
		\leq &\frac{1}{\mu_{\inf}} \sum_{k=0}^{n}\frac{\mu_k^{-1}\beta_kb_k^{2}\phi_k^{\mathrm{T}}P_k\phi_k}{1 + \mu_k^{-1}\beta_k b_k^{2}\phi_k^{\mathrm{T}}P_k\phi_k} \\
        = &O\left(\log\lambda_{\max}(n) \right),
	\end{aligned}
\end{gather}
\noindent where $\mu_{\inf} = \inf\limits_{k\geq0} \mu_k > 0$. The last equality holds due to the boundedness of $\{\mu_k\}_{k\geq 0}$, $\{\beta_k\}_{k\geq 0}$ and $\{b_k\}_{k\geq 0}$.

For the fifth term on the RHS of \eqref{Vc}, by the definition of $w_{k+1}$ in \eqref{wk+1}, we can deduce that $\mathbb{E}_{k}\left[|w_{k+1}|^{\alpha}\right]< \infty$, \text{a.s.}, for any $\alpha \in \left(0,2\right]$. Combing Lemma \ref{marting}, we have
\begin{gather}\label{critical2}
    \begin{aligned}
	&\sum_{k=0}^{n}\mu_k^{-1}\beta_k^{-1}b_k(\psi_k - \beta_k \phi_k^{\mathrm{T}}\tilde{\theta}_k)w_{k+1}\\
        =& o\left( \sum_{k=0}^{n}\mu_k^{-1}\beta_k^{-1}b_k\left(\psi_k-\beta_k \phi_k^{\mathrm{T}}\tilde{\theta}_k\right)^2 \right) + O(1), \quad \text{a.s.}
    \end{aligned}
\end{gather}
\noindent The last equality holds due to the boundedness of $\{\mu_k\}_{k\geq 0}$, $\{\beta_k\}_{k\geq 0}$ and $\{b_k\}_{k\geq 0}$. 

Similarly, we have
\begin{gather}\label{other}
\begin{aligned}
&\sum_{k=0}^{n}\mu_k^{-1}\beta_k^{-1}(1-b_k-\mu_ka_k)\psi_kw_{k+1} \\
=& o\left(\sum_{k=0}^{n} \mu_k^{-1}\beta_k^{-1}b_k\psi_k^2 \right) + O(1), \quad \text{a.s.}
\end{aligned}
\end{gather}

From the definition of $w_{k+1}$ in \eqref{wk+1}, the following fact holds for any $\tau \in (0,2]$,
\begin{gather}
	\sup_{k\geq 0} \mathbb{E}_k \left[ |w_{k+1}^2 - \mathbb{E}_k[w_{k+1}^2]|^{\tau}\right] < \infty,\quad  \text{a.s.}
\end{gather}
Denote $\chi_n = \left(\sum\limits_{k=0}^{n}\left(\mu_k^{-1}a_k\beta_kb_k^2\phi_k^{\mathrm{T}}P_k\phi_k\right)^2\right)^{\frac{1}{2}} $ and let $X_{k}= \sqrt{\mu_{k}^{-1}\beta_k}b_k\phi_k$ in Lemma \ref{lem2}, we have $\chi_n^2 = O(1)$ by noticing the boundedness of $\{a_k\}_{k \geq 0}$, $\{\mu_k\}_{k\geq 0}$, $\{\beta_k\}_{k\geq 0}$ and $\{b_k\}_{k\geq 0}$. Moreover, we have
\begin{gather}\label{33}
	\begin{aligned}
		&\sum_{k=0}^{n}\mu_k^{-1}a_k\beta_kb_k^2 \phi_k^{\mathrm{T}}P_k\phi_k\left(w_{k+1}^2 - \mathbb{E}_k\left[w_{k+1}^2\right]\right)\\
		=&O\left(\chi_n \log^{\frac{1}{2}+\gamma}\left( \chi_n^2 + e\right)\right)\\
		=& O(1), \quad \text{a.s.}, \quad \forall \gamma >0.
	\end{aligned}
\end{gather}

Therefore, from \eqref{step1begin} and \eqref{33}, we know that
\begin{gather}\label{step1end}
	\begin{aligned}
		&\sum_{k=0}^{n}\mu_k^{-1}a_k\beta_kb_k^2 \phi_k^{\mathrm{T}}P_k\phi_k w_{k+1}^2 \\
		\leq  &\sum_{k=0}^{n}\mu_k^{-1}a_k\beta_kb_k^2 \phi_k^{\mathrm{T}}P_k\phi_k\left(w_{k+1}^2 -\mathbb{E}_k\left[ w_{k+1}^2\right] \right) \\
		&+\sup_{k\geq 0}\mathbb{E}_k\left[w_{k+1}^2\right] \left(\sum_{k=0}^{n}\mu_k^{-1}a_k\beta_kb_k^2 \phi_k^{\mathrm{T}}P_k\phi_k\right)\\
		=&O(\log \lambda_{\max}(n)), \quad \text{a.s.}
	\end{aligned}
\end{gather}

The analysis of the third term on the RHS of (\ref{Vc}) is critical for the concatenation of the two step of the TSWLAD algorithm.  Before analyzing it, we introduce the following fact based on Lemma \ref{critical}, and noticing the boundedness of $\{\mu_k\}_{k \geq 0}$, $\{\beta_k\}_{k \geq 0}$ and $\{b_k\}_{k \geq 0}$:
\begin{gather}\label{trans}
\sum\limits_{k=0}^{n}\mu_k^{-1}\beta_k^{-1}b_k( \psi_k - \beta_k\phi_k^{\mathrm{T}}\tilde{\theta}_k)^2= O(\sum\limits_{k=0}^{n}(\phi_k^{\mathrm{T}}\tilde{\bar{\theta}}_k)^2),\quad \text{a.s.} 
\end{gather}

We are now in position to prove 
\begin{gather}\label{step2}
	\sum_{k=0}^{n} \left( \phi_{k+1}^{\mathrm{T}}\tilde{\bar{\theta}}_{k+1}\right)^2=O\left(\log\lambda_{\max}(n)\right), \quad \text{a.s.}
\end{gather}
To achieve this, we  consider the following similar Lyapunov function:
\begin{gather}
	\bar{V}_{k+1} = \tilde{\bar{\theta}}_{k+1}^{\mathrm{T}}\bar{P}_{k+1}^{-1}\tilde{\bar{\theta}}_{k+1} ,
\end{gather}
by Lemma \ref{lemma3}, we know that $\phi_k^{\mathrm{T}}\tilde{\bar{\theta}}_k\bar{\psi}_k\geq 0$ and $|\bar{\psi}_k| \geq \bar{\beta}_k |\phi_k^{\mathrm{T}}\tilde{\bar{\theta}}_k|$, and noticing that $0<b_k\leq 1$, we have 
\begin{gather}\label{f1}
\begin{aligned}
&\bar{\mu}_k^{-1}\bar{\beta}_kb_k^2\left(\phi_k^{\mathrm{T}}\tilde{\bar{\theta}}_k \right)^2 - 2\bar{\mu}_k^{-1}b_k\phi_k^{\mathrm{T}}\tilde{\bar{\theta}}_k\bar{\psi}_k \\
=&\bar{\mu}_k^{-1}\bar{\beta}_kb_k^2\left(\phi_k^{\mathrm{T}}\tilde{\bar{\theta}}_k \right)^2-2\bar{\mu}_k^{-1}b_k|\phi_k^{\mathrm{T}}\tilde{\bar{\theta}}_k| |\bar{\psi}_k| \\
\leq& -\bar{\mu}_k^{-1}\bar{\beta}_kb_k\left(\phi_k^{\mathrm{T}}\tilde{\bar{\theta}}_k \right)^2.
\end{aligned}
\end{gather}

Following similar analysis of (\ref{Vc}) and combing \eqref{f1}, we have
\begin{gather}\label{connection}
	\begin{aligned}
		\bar{V}_{n+1} \leq& \bar{V}_0 - \sum_{k=0}^{n} \bar{\mu}_k^{-1}\bar{\beta}_kb_k\left(\phi_k^{\mathrm{T}}\tilde{\bar{\theta}}_k \right)^2  \\
		& +\frac{1}{h}\sum_{k=0}^{n} \bar{\mu}_k^{-1}\bar{a}_k\bar{\beta}_kb_k^2\phi_k^{\mathrm{T}}\bar{P}_k\phi_k \\
		&+2\sum_{k=0}^{n}\bar{\mu}_k^{-1}\bar{a}_kb_k^2\phi_k^{\mathrm{T}}\bar{P}_k\phi_k\bar{\psi}_k\bar{w}_{k+1}\\
            &  -2\sum_{k=0}^{n}\bar{\mu}_k^{-1}b_k\phi_k^{\mathrm{T}}\tilde{\bar{\theta}}_k\bar{w}_{k+1} \\
             &+ \frac{1}{h}\sum_{k=0}^{n} \bar{\mu}_k^{-1}\bar{a}_k\bar{\beta}_k b_k^2\phi_k^{\mathrm{T}}\bar{P}_k\phi_k\bar{w}_{k+1}^2, \quad \text{a.s.},
	\end{aligned}
\end{gather}

\noindent where $h$ is specified in \eqref{Vc}.

Drawing upon similar analysis in \eqref{step1begin}--\eqref{step1end} to deal with \eqref{connection}, we can derive the following property:
\begin{gather}\label{step1critical}
	\bar{V}_{n+1} + \sum_{k=0}^{n}\bar{\mu}_k^{-1}\bar{\beta}_kb_k\left(\phi_k^{\mathrm{T}}\tilde{\bar{\theta}}_k \right)^2 = O(\log \lambda_{\max}(n)), \quad \text{a.s.}
\end{gather}
Since $\left\{\bar{\mu}_k\right\}_{k \geq 0},\left\{\bar{\beta}_k\right\}_{k \geq 0}$ and $\left\{b_k\right\}_{k \geq 0}$ are bounded sequences almost surely, it can be concluded that (\ref{step2}) holds based on (\ref{step1critical}).  

Ultimately, with \eqref{Vc}, \eqref{step1begin}, \eqref{critical2}, \eqref{other}, \eqref{step1end}, \eqref{trans} and \eqref{step2}, we get the desired result \eqref{desired}. $\hfill\blacksquare$

\noindent\hspace{1em}{\textbf{\itshape Proof of Theorem \ref{Th1}:}}  With Lemma \ref{lemma1}, \eqref{Vk} and \eqref{Pni}, we can have that
\begin{gather}\label{proofTh1}
    V_{n+1} \geq c_0\lambda_{\min}(k)\| \tilde{\theta}_{n+1}\|^2, ~\text{a.s.},
\end{gather}
\noindent where $c_0=\min\{1, \inf\limits_{k\geq 0}(\mu_k^{-1}\beta_kb_k^2)\}$ is positive since the infimum of $\{\mu_k\}_{k \geq 0}$, $\{b_k\}_{k \geq 0}$ and $\{\beta_k\}_{k \geq 0}$ are greater than 0. Therefore, Theorem \ref{Th1} follows immediately from Lemma \ref{lemma1} and \eqref{proofTh1}. $\hfill\blacksquare$

\noindent\hspace{1em}{ \textbf{\itshape Proof of Theorem \ref{Th2} and Theorem \ref{Th3}:}}  By Lemma \ref{lemma3} and the boundedness condition \eqref{boundednessf}, we have that
\begin{gather}\label{proofTh2}
    \psi_k^2 \geq h^2(\phi_k^{\mathrm{T}}\tilde{\theta}_k)^2,
\end{gather}
\noindent where $h$  defined in \eqref{Vc} is positive. By Lemma \ref{lemma1} and combing the boundedness of $\{\mu_k\}_{k\geq 0}$, $\{\beta_k\}_{k\geq 0}$ and $\{b_k\}_{k\geq 0}$, we have that 
\begin{gather} \label{ptsquare}
    \sum_{k=1}^{n}b_k^2 (\phi_k^{\mathrm{T}}\tilde{\theta}_k)^2 = O(\log\lambda_{\max}(n) ), \quad\text{a.s.},
\end{gather}

Since $l_k= L_k$ and $u_k=U_k$, we can deduce that $|S_k(x) - S_k(y)| \leq |x-y|$  for any $x\in \mathbb{R}$, $y\in \mathbb{R}$, then by the Cauchy-Schwarz inequality, we have 
\begin{gather}
\sum_{k=1}^{n}R_k \leq \sqrt{n \sum_{k=1}^{n}b_k^2 (\phi_k^{\mathrm{T}}\tilde{\theta}_k)^2} =O(\sqrt{n\log\lambda_{\max}(n)}),\ \text{a.s.},
\end{gather}
\noindent then Theorem \ref{Th2} follows. 

For the proof of Theorem \ref{Th3}, we have 
\begin{gather}\label{preerror1}
\begin{aligned}
    &\sum_{k=0}^{n-1} b_k^2(y_{k+1} - \hat{y}_{k+1})^2 \\
    =& \sum_{k=0}^{n-1} b_k^2(S_k(\phi_k^{\mathrm{T}}\theta +\varepsilon_{k+1}) -S_k(\phi_k^{\mathrm{T}}\theta_k))^2 \\
    \leq& \sum_{k=0}^{n-1}b_k^2(\phi_k^{\mathrm{T}}\tilde{\theta}_k + \varepsilon_{k+1})^2 \\
     =& \sum_{k=0}^{n-1}b_k^2(\phi_k^{\mathrm{T}}\tilde{\theta}_k)^2 + 2\sum_{k=0}^{n-1}b_k^2(\phi_k^{\mathrm{T}}\tilde{\theta}_k)\varepsilon_{k+1}+\sum_{k=0}^{n-1}b_k^2\varepsilon_{k+1}^2,
\end{aligned}
\end{gather}
\noindent we now analyze the RHS of \eqref{preerror1} term  by term. First, according to \eqref{ptsquare}, we can deduce that the first term on the RHS of \eqref{preerror1} is bounded by $O(\log\lambda_{\max}(n))$. By Lemma \ref{marting} and noticing the boundedness of $b_k$, we have
\begin{gather}\label{pte}
\begin{aligned}
    &\sum_{k=0}^{n-1}b_k^2(\phi_k^{\mathrm{T}}\tilde{\theta}_k)\varepsilon_{k+1} \\
    =&O\left( \left\{\sum_{k=0}^{n-1}b_k^2(\phi_k^{\mathrm{T}}\tilde{\theta}_k)^2\right\}^{\frac{1}{2}+\eta}\right) \\
    = & o\left(\sum_{k=0}^{n-1}b_k^2(\phi_k^{\mathrm{T}}\tilde{\theta}_k)^2 \right) + O(1), \quad\text{a.s.}, \quad \forall \eta > 0,
\end{aligned}
\end{gather}
\noindent since $\{\varepsilon_{k}, \mathcal{F}_k\}_{k \geq 0}$ is a martingale difference sequence and satisfies \eqref{noisemarting}.

Moreover, by \eqref{noisemarting}, the $C_r$-inequality and the Lyapunov inequality, we have
\begin{gather}
\begin{aligned}
    &\sup_{k\geq 0} \mathbb{E}_k\left\{\left|\varepsilon_{k+1}^2 - \mathbb{E}_k\left[\varepsilon_{k+1}^2 \right]\right|^{\frac{r}{2}}\right\} \\
    &\leq 2^{\frac{r}{2}} \sup_{k \geq 0} \mathbb{E}_{k}\left[\left|\varepsilon_{k+1}\right|^{r} \right] < \infty, \quad \text{a.s.},
\end{aligned}   
\end{gather}
therefore, by a refined martingale estimation theorem (see \cite{guo2020tima}) and noticing the boundedness of $b_k$, we can deduce that 
\begin{gather} \label{e2c}
    \sum_{k=0}^{n-1} b_k^2\left\{\varepsilon_{k+1}^2 - \mathbb{E}_k\left[\varepsilon_{k+1}^2 \right] \right\} = O\left( \sqrt{n\log\log n}\right), \ \text{a.s.}
\end{gather}

Combing \eqref{preerror1}, \eqref{pte} and \eqref{e2c}, we have
\begin{gather} 
\begin{aligned}
    &\sum_{k=0}^{n-1} b_k^2(y_{k+1} - \hat{y}_{k+1})^2 \\
    &=O\left(\log\lambda_{\max}(n) + \sqrt{n\log\log n}\right) + \sum_{k=0}^{n-1}b_k^2\sigma_k^2, \quad \text{a.s.},   
\end{aligned}
\end{gather}
therefore, by using the Cauchy-Schwarz inequality and the boundedness of $b_k$, and noticing that $\sqrt{x+y}\leq \sqrt{x}+\sqrt{y} $ for any $x\geq 0$ and $y \geq 0$, we know that the following fact holds almost surely:
\begin{gather}
\begin{aligned}
    E_n &\leq \frac{1}{n} \sqrt{n}\sqrt{\sum_{k=0}^{n-1}b_k^2(y_{k+1} - \hat{y}_{k+1})^2}\\
    &\leq  \sqrt{\frac{\sum\limits_{k=0}^{n-1}b_k^2\sigma_k^2}{n}} + O\left( \sqrt{\frac{\log\lambda_{\max}(n)}{n} + \sqrt{\frac{\log\log n}{n}}  }\right) ,
\end{aligned}
\end{gather}
this complete the proof of Theorem \ref{Th3}. $\hfill\blacksquare$

\section{Simulation and Experiment} \label{simulation}

In this section, we will demonstrate the advantages of the new $\ell_1$-optimization-based adaptive algorithm (TSWLAD) over the previous $\ell_2$-optimization-based algorithm (TSQN) proposed in \cite{zhang2023adaptive} via both numerical examples and real-data-based experiments.

\subsection{Numerical example}
Considering the following saturated stochastic system:
\begin{gather} \label{simulationexample}
    \left\{ \begin{array}{cl}
    \phi_{k+1} &= A\phi_k + v_{k+1},\\
    y_{k+1}&= S_k(\phi_k^{\mathrm{T}}\theta + \varepsilon_{k+1}),
    \end{array} \right.
\end{gather}
\noindent where the saturated function $S_k(\cdot)$ is specified as follows:
\begin{gather} 
	S_k(x)=\left\{\begin{array}{cl}
	0, & x\leq0, \\
	x, & 0 < x < 25 , \\
	25, & x \geq 25,
	\end{array}\right.
\end{gather} 
\noindent and where $A= \operatorname{diag}[0.99, 0.5, 0.9, 0.01, 0.3, 0.7]$, $\operatorname{diag}[\cdot]$ denotes the diagonal matrix,  the true parameter vector is $\theta = [5, 0.7, 2, -0.1, -0.6, -8]^{\mathrm{T}}$, the noise $v_{k+1} = (v_{k+1}^{(1)}, \cdots, v_{k+1}^{(6)})^{\mathrm{T}}$  in the state equation satisfies the condition that $v_{k+1}^{(i)}$ are independent with $v_{k+1}^{(1)} \sim N(0, 1)$ and $v_{k+1}^{(j)}\sim \frac{5}{\sqrt[4]{k+1}}N(0, 1)$ for any $k \geq 0$ and $j=2, \cdots, 6$, $\phi_k \in \mathbb{R}^{6}$, $\phi_0 = \mathbf{0}$, it is not difficult to verify that the regression vectors generated by \eqref{simulationexample} do not satisfy the traditional PE condition, but do satisfy the much weaker eigenvalue condition \eqref{npe}, which can guarantee the convergence of the TSWLAD and TSQN algorithms. Usually, one may assume that the noise $\varepsilon_{k+1}$ is independent and follows the distribution $N(0,1)$, however, the distribution of the noise $\varepsilon_{k+1}$ may switch to $N(0,10)$ with a certain probability due to unexpected disturbances (such as sensor malfunctions). So we assume that  the noise $\varepsilon_{k+1}$ following either the standard normal distribution $N(0,1)$ with probability $1 - q$, or the normal distribution $N(0,10)$ with probability $q$, where $0 \leq q \leq 1$. 

To show the advantages of our newly proposed TSWLAD algorithm, we will compare the parameter estimation errors of the TSWLAD algorithm with that of the TSQN algorithm based on a dataset $\{\phi_i, y_{i+1}\}_{i=1}^{10000}$ generated from the system \eqref{simulationexample} under different $q$ values. In the process of computation, the TSWLAD algorithm adopts the following configuration: the convex compact set $D=\{x \in \mathbb{R}^6:|x^{(i)}| \leq 10\}$, $b_k = 1$, $\bar{\mu}_k=\mu_k=1$, $\bar{\theta}_0=\theta_0 =\mathbf{0}$, $\bar{P}_0 = P_0 = I$; Assuming that we are unaware of the presence of the unexpected disturbances, the corresponding conditional density function $f_{k+1}(\cdot)$ and  distribution function $F_{k+1}(\cdot)$ are chosen as those of the normal distribution $N(0, 1)$, respectively. The TSQN algorithm is configured in the same way for fair comparison of the estimation errors.

\begin{figure}
  \centering
  \includegraphics[width= 0.826\linewidth]{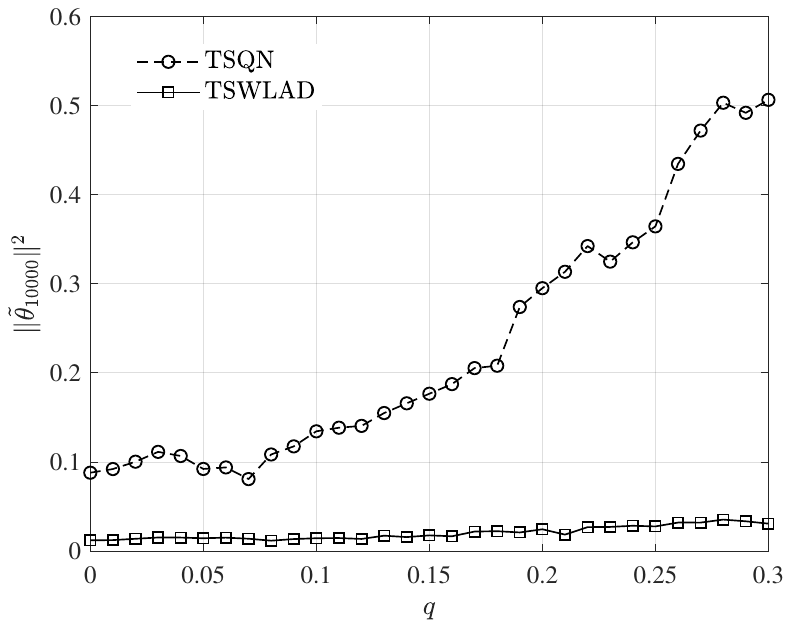}
  \caption{Comparison of parameter estimation errors for the two algorithms under different $q$ values, \(\tilde{\theta}_{10000}\) is the parameter estimation error computed at time $k=10000$.}
  \label{figurec}
\end{figure}

\begin{table}[t]
\caption{Comparison of Parameter Estimation Errors $\tilde{\theta}_{10000}$ under Some $q$ Values }
\label{tablec}
\begin{center}
\resizebox{0.37\textwidth}{!}{
\begin{tabular}{ccccc}
\hline
  $q$ &0 & 0.1 & 0.2  & 0.3  \\ \hline
TSQN & 0.0880 & 0.1345 & 0.2952 & 0.5067 \\
TSWLAD & 0.0121 & 0.0144  &0.0245 & 0.0308\\
\hline
\end{tabular}}
\end{center}
\end{table}

Both Fig. \ref{figurec}  and Table \ref{tablec} demonstrate that the TSWALD algorithm is more robust in adaptive identification compared to the TSQN algorithm (\cite{zhang2023adaptive}) across various scenarios involving unexpected disturbances affecting the system observations. 

The following Fig. \ref{figure1} and Fig. \ref{figure2} show the averaged regrets of the corresponding adaptive predictors of the TSWLAD algorithm with $q=0$.

\begin{figure}
  \centering
  \includegraphics[width= 0.83\linewidth]{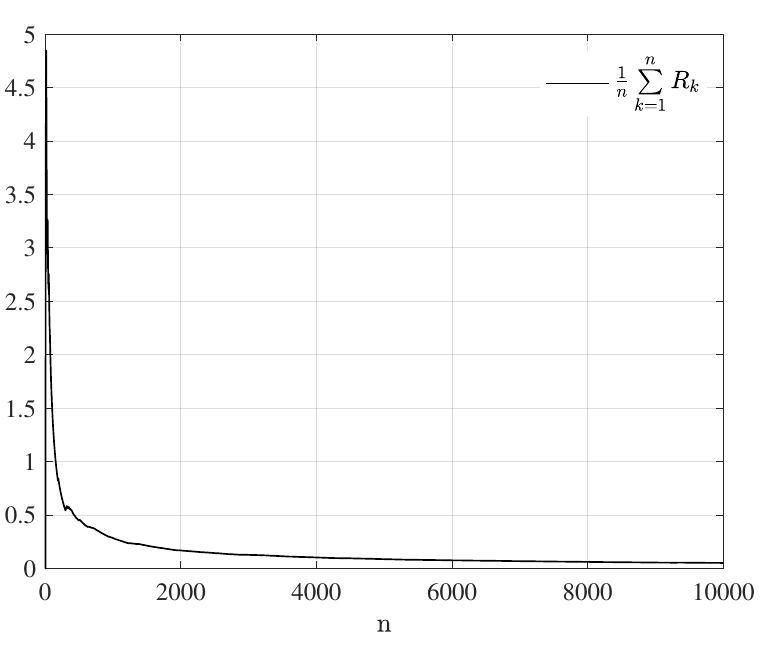}
  \caption{A trajectory of $\frac{1}{n}\sum\limits_{k=1}^{n}R_k$.}
  \label{figure1}
\end{figure}

\begin{figure} 
  \centering
  \includegraphics[width= 0.83\linewidth]{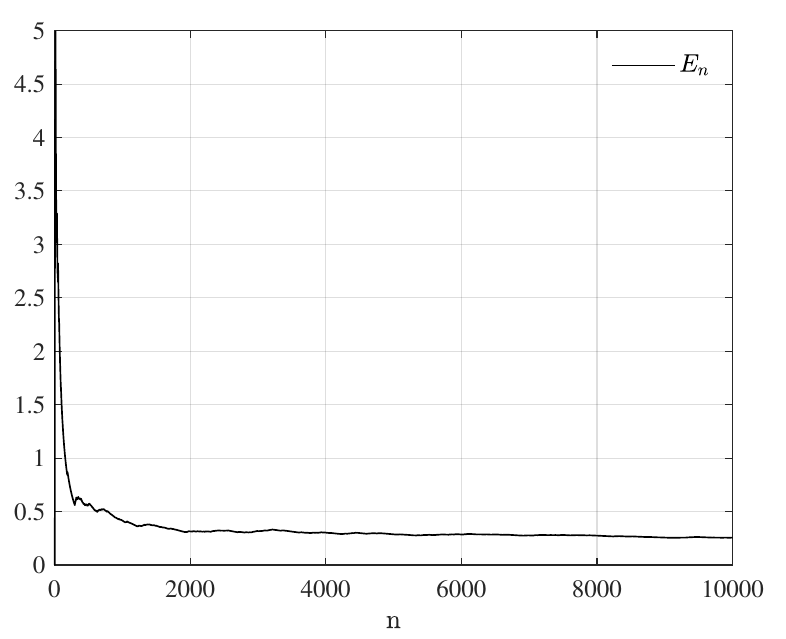}
  \caption{A trajectory of $E_n$.}
  \label{figure2}
\end{figure}

\subsection{Real-data-based judicial sentencing problem}
 With the rapid development of information technology, data science plays an increasingly important role in the construction of smart court in China. The application of judicial data contributes to the establishment of intelligent adjudication assistant systems within smart courts, thereby assisting in judicial sentencing. Given that sentencing directly relates to fairness and justice, there is significant public concern regarding its normative aspects and fairness, leading to high demands for the interpretability of the computational results derived from judicial data \cite{wang2022applications}. Adaptive identification techniques have demonstrated remarkable capabilities in processing judicial data, as they impose relatively lenient requirements on data quality (not necessary iid data) while maintaining high reliability in theory. According to the related laws in China, the announced penalties of criminals are constrained to some statutory ranges depending on criminal plots, renders judicial sentencing data as saturated output observations, and the related data may be obtained from the China Judgments Online.  We now use part of the intentional injury crime sentencing data to assess the adaptive predictive performance of the TSWLAD algorithm and compare it with the existing TSQN algorithm. The judicial sentencing data spanning the years 2011 to 2021. Within this dataset, there are 171,666 cases involving only minor injuries and 24,503 cases involving serious injuries. Based on the dataset, we conducted experiments with the TSWLAD algorithm and the TSQN algorithm for prediction accuracy comparisons, the evaluation metric is as follows:  
\begin{gather}\label{sentencing_loss}
    1 - \frac{1}{T}\sum_{k=1}^{T}\frac{|y_k - \hat{y}_k|}{y_k},
\end{gather}
\noindent where \( T \) denotes the total number of sentences adjudicated.

In the process of computation, the TSWLAD algorithm adopts the the following configuration: the noise assumption is the same as that used in \cite{wang2022applications}, i.e., $\{\varepsilon_k\}$ are independent and follow the distribution $\mathrm{N}(0, 25)$, the convex compact set \(D\) is the same as in \cite{wang2022applications}, \(C_k = \sup\limits_{x \in D} |\phi_k^{\mathrm{T}} x|\), $\bar{\mu}_k=1000$, $\mu_k=25$, $b_k=\frac{1}{\hat{y}_k}$ and the initial matrices are set as $\bar{P}_0 = P_0 = I$, the initial parameters $\bar{\theta}_0$ and $\theta_0$ are estimated based on the least squares method. The hyperparameters of the TSQN algorithm is the same as that in \cite{wang2022applications}.

\begin{figure}[ht] 
  \centering
  \includegraphics[width= 0.84\linewidth]{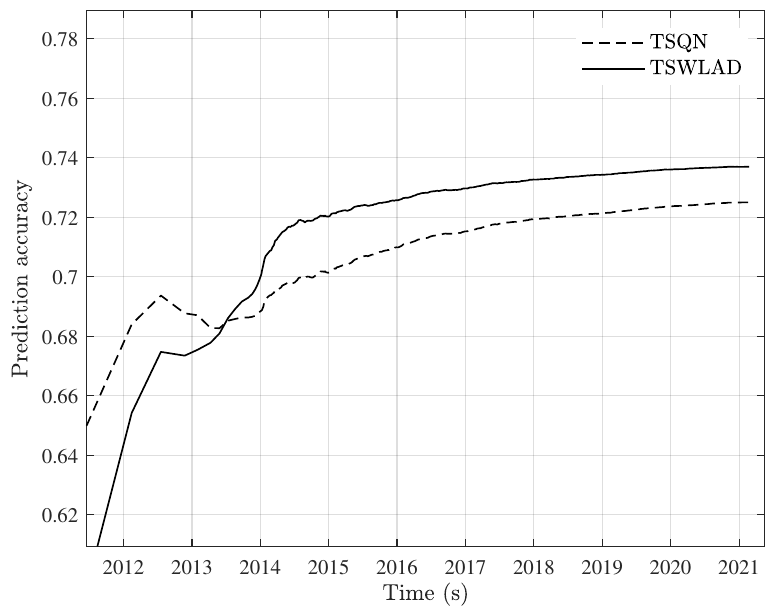}
  \caption{Comparison of adaptive prediction accuracy in minor injury cases.}
  \label{qing}
\end{figure}

\begin{figure}[ht] 
  \centering
  \includegraphics[width= 0.84\linewidth]{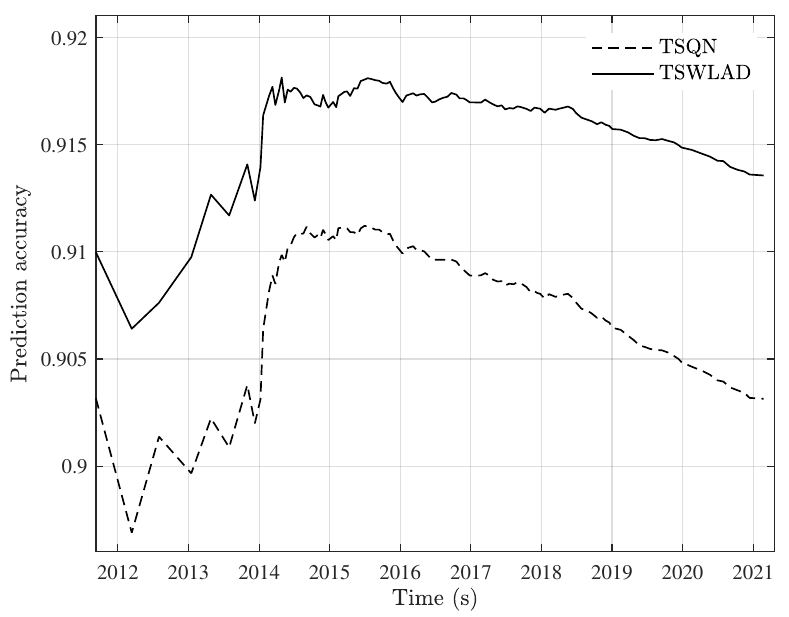}
  \caption{Comparison of adaptive prediction accuracy in serious injury cases.}
  \label{zhong}
\end{figure}

Based on the collected data,  Fig. \ref{qing} and Fig. \ref{zhong} show that the TSWLAD algorithm outperforms the TSQN algorithm in terms of sentencing prediction accuracy for both minor injury and serious injury cases.

\section{Conclusion}\label{conclusion}

Inspired by various real-world application problems, we have proposed a new nonlinear adaptive identification algorithm based on both saturated observations and the weighted $\ell_1$-norm optimization. Under quite general non-PE conditions on system signals that may come from feedback control systems, we have established the global convergence of the parameter estimates and the upper bounds of the averaged regrets of adaptive predictions.
The superior robustness and effectiveness of the proposed $\ell_1$-norm based  adaptive algorithms are demonstrated in comparison with the existing $\ell_2$-norm based adaptive algorithms, via a simulation example and a real application scenario of judicial sentencing data analysis and prediction. For future investigations, the carefully selected weighting coefficients $\{b_k\}$ may be further explored to enhance the robustness of the adaptive algorithm. Moreover, different adaptive algorithms for different saturated nonlinear models may also be studied to meet with the needs in different application scenarios.

\appendices
\section{}\label{app1}
\noindent\hspace{1em}{\textbf{\itshape Proof of Lemma \ref{l1}:}} 
We  only elaborate the proof of $|\bar{\psi}_k| \leq 1$ since the proof of $|\psi_k| \leq 1$ follows the same way. In this proof, \(\mathbb{P}_{k+1}(\cdot)\) represents the conditional probability function of \(y_{k+1}\) given the $\sigma$-algebra $\mathcal{F}_k$. The proof of $|\bar{\psi}_k| \leq 1$ is established in three distinct cases below.

If $S_k(\phi_k^{\mathrm{T}}\bar{\theta}_{k})=L_k$, noticing that $\mathbb{P}_{k+1}(y_{k+1}< L_k)=0$, we have
\begin{equation}
    \begin{aligned}
	\bar{\psi}_{k} =& \mathbb{E}_k\left\{\operatorname{sgn}\left[y_{k+1}-L_k\right]\right\}-\left[1-F_{k+1}(l_k-\phi_k^{\mathrm{T}}\bar{\theta}_{k})\right]\\
	=&\mathbb{P}_{k+1}(y_{k+1}> L_k)-\left[1-F_{k+1}(l_k-\phi_k^{\mathrm{T}}\bar{\theta}_{k})\right]	\\
	=&1-\mathbb{P}_{k+1}\left(S_k(\phi_k^{\mathrm{T}}\theta+\varepsilon_{k+1})\leq L_k\right)\\
         &-\left[1 - F_{k+1}(l_k-\phi_k^{\mathrm{T}}\bar{\theta}_k)\right]		\\
	 =& F_{k+1}(l_k - \phi_k^{\mathrm{T}}\bar{\theta}_{k})-F_{k+1}(l_k-\phi_k^{\mathrm{T}}\theta),		
    \end{aligned}
\end{equation}
\noindent where the last equality holds since the density function $f_{k+1}(\cdot)$ is continuous.
 
If $L_k < S_k(\phi_k^{\mathrm{T}}\bar{\theta}_{k}) < U_k$, then $S_k(\phi_k^{\mathrm{T}}\bar{\theta}_{k})=\phi_k^{\mathrm{T}}\bar{\theta}_k$, we have
\begin{equation}
    \begin{aligned}
	\bar{\psi}_{k}  =& \mathbb{E}_k\left\{\operatorname{sgn}\left[y_{k+1} - \ptb \right]\right\}\\
	=& \mathbb{P}_{k+1}\left(y_{k+1}>\phi_k^{\mathrm{T}}\bar{\theta}_{k}\right) - \mathbb{P}_{k+1}\left(y_{k+1}< \phi_k^{\mathrm{T}}\bar{\theta}_{k}\right)\\
	=& 1 - \mathbb{P}_{k+1}\left(S_k(\phi_k^{\mathrm{T}}\theta+\varepsilon_{k+1})\leq \ptb\right)\\
          &- \mathbb{P}_{k+1}\left(S_k(\phi_k^{\mathrm{T}}\theta+\varepsilon_{k+1})< \ptb\right)\\
	 =& 1 - 2F_{k+1}(\phi_k^{\mathrm{T}}\bar{\theta}_{k} - \phi_k^{\mathrm{T}}\theta),
    \end{aligned}
\end{equation}
 \noindent where the last equality holds because of the continuity of the density function $f_{k+1}(\cdot)$.
   
If $S_k(\phi_k^{\mathrm{T}}\bar{\theta}_{k})=U_k$, noticing that  $P_k(y_{k+1} > U_k)=0$, we know that
\begin{equation}
    \begin{aligned}
	\bar{\psi}_k &=  \mathbb{E}_k\left\{\operatorname{sgn}\left[y_{k+1}-U_k\right]\right\} + F_{k+1}\left(u_k-\phi_k^{\mathrm{T}}\bar{\theta}_{k}\right)  \\
	&= -\mathbb{P}_{k+1}\left(y_{k+1} <U_k\right)+F_{k+1}\left(u_k-\phi_k^{\mathrm{T}}\bar{\theta}_{k}\right)		\\
	  &=F_{k+1}\left(u_k - \phi_k^{\mathrm{T}}\bar{\theta}_{k}\right) -\mathbb{P}_{k+1}\left(S_k(\phi_k^{\mathrm{T}}\theta+e_{k+1})<U_k\right) \\
	  &= F_{k+1}\left(u_k-\phi_k^{\mathrm{T}}\bar{\theta}_{k}\right)-F_{k+1}\left(u_k-\phi_k^{\mathrm{T}}\theta\right),
    \end{aligned}
\end{equation}
\noindent where the last equality holds due to the continuity of the density function $f_{k+1}(\cdot)$.

Thus far, all of the above cases indicate that $|\bar{\psi}_k |\leq 1$. $\hfill\blacksquare$

\noindent\hspace{1em}{\textbf{\itshape Proof of Lemma \ref{critical}:}} This proof is completed under three different scenarios below.

If $S_k\left(\phi_k^{\mathrm{T}}\theta_{k}\right) = L_k$, we have 
\begin{equation}
    \begin{aligned}
	&\psi_k - \left[F_{k+1}(l_k - \pt)-F_{k+1}(l_k - \ptb)\right]\\ 
	=&\psi_k - \beta_k \phi_k^{\mathrm{T}}(\bar{\theta}_k - \theta_k) \\
	=& \psi_k - \beta_k \phi_k^{\mathrm{T}}\tilde{\theta}_k + \beta_k \phi_k^{\mathrm{T}}\tilde{\bar{\theta}}_k,
    \end{aligned}
\end{equation}
\noindent since $\psi_k = F_{k+1}(l_k - \pt) - F_{k+1}(l_k - \phi_k^{\mathrm{T}}\theta) $, then $\psi_k - \beta_k \phi_k^{\mathrm{T}}\tilde{\theta}_k = \left(f_{k+1}(z_1)- \beta_k\right)\phi_k^{\mathrm{T}}\tilde{\bar{\theta}}_k$, where the value $z_1$ is between $ l_k - \ptb$ and $l_k - \phi_k^{\mathrm{T}}\theta$ according to the mean value theorem.	

If $ L_k < S_k\left(\phi_k^{\mathrm{T}}\theta_{k}\right)< U_k$, the following fact holds:	
\begin{gather}
    \begin{aligned}
        &\psi_k - [1 - 2F_{k+1}(\pt - \ptb)] \\
	 =&\psi_k - \beta_k \phi_k^{\mathrm{T}}(\bar{\theta}_k - \theta_k) \\
	=& \psi_k - \beta_k \phi_k^{\mathrm{T}}\tilde{\theta}_k + \beta_k \phi_k^{\mathrm{T}}\tilde{\bar{\theta}}_k,
    \end{aligned}
\end{gather}
\noindent since $\psi_k = 1 - 2F_{k+1}(\pt - \phi_k^{\mathrm{T}}\theta)$, then $\psi_k -\beta_k\phi_k^{\mathrm{T}}\tilde{\theta}_k = \left[2f_{k+1}(z_2) - \beta_k\right]\phi_k^{\mathrm{T}}\tilde{\bar{\theta}}_k$, where the value $z_2$ takes values between $\phi_k^{\mathrm{T}}(\theta_k - \bar{\theta}_k) $ and $-\phi_k^{\mathrm{T}}\tilde{\theta}_{k}$ according to the mean value theorem.
	
If $S_k\left(\phi_k^{\mathrm{T}}\theta_{k}\right) = U_k$, we know that
\begin{gather}
    \begin{aligned}
	&\psi_k  -[F_{k+1}(u_k - \pt) - F_{k+1}(u_k - \ptb)]\\
	=& \psi_k - \beta_k\phi_k^{\mathrm{T}}(\bar{\theta}_k - \theta_k) \\
	=& \psi_k - \beta_k \phi_k^{\mathrm{T}}\tilde{\theta}_k + \beta_k \phi_k^{\mathrm{T}}\tilde{\bar{\theta}}_k,
    \end{aligned}
\end{gather}
\noindent since $\psi_k = F_{k+1}(u_k - \pt) - F_{k+1}(u_k - \phi_k^{\mathrm{T}}\theta) $, then $\psi_k - \beta_k \phi_k^{\mathrm{T}}\tilde{\theta}_k = \left(f_{k+1}(z_3) - \beta_k\right)\phi_k^{\mathrm{T}}\tilde{\bar{\theta}}_k$, where the $z_3$ takes values between $ u_k - \ptb$ and $u_k - \phi_k^{\mathrm{T}}\theta$.
	
Under Assumption \ref{symmetry}, we know that $f_{k+1}(z_i) (i=1,2,3)$ and $\beta_k$ are bounded almost surely. Hence Lemma \ref{critical} holds. $\hfill\blacksquare$

\noindent\hspace{1em}{\textbf{\itshape Proof of Lemma \ref{lemma3}:}} This proof is completed under three different cases below.

If $S_k\left(\phi_k^{\mathrm{T}}\theta_{k}\right) = L_k$, we know that
\begin{equation}
    \begin{aligned}
	\bar{\psi}_{k} &= F_{k+1}(l_k - \phi_k^{\mathrm{T}}\bar{\theta}_{k})-F_{k+1}(l_k-\phi_k^{\mathrm{T}}\theta)	\\	
      &= f_{k+1}(e_1)\phi_k^{\mathrm{T}}\tilde{\bar{\theta}}_k,
    \end{aligned}
\end{equation}
\noindent where the $e_1$ is between $ l_k - \ptb$ and $l_k - \phi_k^{\mathrm{T}}\theta$ according to the mean value theorem. Under Assumption \ref{symmetry}, we know that $f_{k+1}(e_1)\geq \bar{\beta}_k$, hence $|\bar{\psi}_k| \geq \bar{\beta}_k |\phi_k^{\mathrm{T}}\tilde{\bar{\theta}}_k|$. Besides, $\phi_k^{\mathrm{T}}\tilde{\bar{\theta}}_k\bar{\psi}_k = f_{k+1}(e_1) (\tilde{\bar{\theta}}_k\bar{\psi}_k)^2\geq 0$.

If $L_k <S_k\left(\phi_k^{\mathrm{T}}\theta_{k}\right) < U_k$, combing  Assumption \ref{symmetry}, we know that
\begin{equation}
    \begin{aligned}
	\bar{\psi}_{k} &= 1 - 2F_{k+1}(\phi_k^{\mathrm{T}}\bar{\theta}_{k} - \phi_k^{\mathrm{T}}\theta)	\\	
     & =2F_{k+1}(0)-2F_{k+1}(\phi_k^{\mathrm{T}}\bar{\theta}_{k} - \phi_k^{\mathrm{T}}\theta)\\
      &= 2f_{k+1}(e_2)\phi_k^{\mathrm{T}}\tilde{\bar{\theta}}_{k},
    \end{aligned}
\end{equation}
\noindent where the $e_2$ is between 0 and $-\phi_k^{\mathrm{T}}\tilde{\bar{\theta}}_k$ according to the mean value theorem. Under  Assumption \ref{symmetry}, we know that $f_{k+1}(e_2)\geq \bar{\beta}_k $, hence $|\bar{\psi}_k| \geq \bar{\beta}_k |\phi_k^{\mathrm{T}}\tilde{\bar{\theta}}_k|$. Besides, $\phi_k^{\mathrm{T}}\tilde{\bar{\theta}}_k\bar{\psi}_k = 2f_{k+1}(e_2) (\tilde{\bar{\theta}}_k\bar{\psi}_k)^2\geq 0$.

As for the case where $S_k(\phi_k^{\mathrm{T}}\bar{\theta}_{k})=U_k$, we know that
\begin{equation}
    \begin{aligned}
	\bar{\psi}_{k} &= F_{k+1}(u_k - \phi_k^{\mathrm{T}}\bar{\theta}_{k})-F_{k+1}(u_k-\phi_k^{\mathrm{T}}\theta)	\\	
      &= f_{k+1}(e_3)\phi_k^{\mathrm{T}}\tilde{\bar{\theta}}_k,
    \end{aligned}
\end{equation}
\noindent where the $e_3$ is between $ u_k - \ptb$ and $u_k - \phi_k^{\mathrm{T}}\theta$ according to the mean value theorem. Under Assumption \ref{symmetry}, we have $f_{k+1}(e_3)\geq \bar{\beta}_k$, hence $|\bar{\psi}_k| \geq \bar{\beta}_k |\phi_k^{\mathrm{T}}\tilde{\bar{\theta}}_k|$. Besides, $\phi_k^{\mathrm{T}}\tilde{\bar{\theta}}_k\bar{\psi}_k = f_{k+1}(e_3) (\tilde{\bar{\theta}}_k\bar{\psi}_k)^2\geq 0$.

Then this Lemma holds. $\hfill\blacksquare$

\bibliographystyle{IEEEtran}
\bibliography{root}

\end{document}